\documentstyle[aps,preprint]{revtex}
\begin{document}
\title{
Superconductivity in quantum-dot superlattices composed of quantum wire networks
}
\author{
Takashi~Kimura${}^1$, Hiroyuki~Tamura${}^1$, Kazuhiko~Kuroki${}^2$,
 Kenji~Shiraishi${}^{3,4}$, Hideaki~Takayanagi${}^1$, and Ryotaro~Arita${}^5$
}
\address{
${}^1$NTT Basic Research Laboratories, 
NTT Corporation, Atsugi 243-0198, Japan \\ 
${}^2$Department of Applied Physics and Chemistry, 
The University of Electro-Communications, Chofu 182-8585, Japan \\
${}^3$Department of Physics, Tsukuba University, Tsukuba 305-8271, Japan \\
${}^4$Research Consortium for Synthetic Nano-Function Materials Project (SYNAF), 
National Institute of Advanced Industrial Science and Technology (AIST), 
Tsukuba 305-8568, Japan \\
${}^5$ Department of Physics, University of Tokyo, Hongo 113-0033, Japan
}
\date{\today}
\maketitle
\begin{abstract}
Based on calculations using the local density approximation, we propose quantum wire networks with square and plaquette type lattice structures that form quantum dot superlattices. These artificial structures are well described by the Hubbard model. Numerical analysis reveals a superconducting ground state with transition temperatures $T_c$ of up to 90 mK for the plaquette, which is more than double the value of 40 mK for the square lattice type and is sufficiently high to allow for the experimental observation of superconductivity.
\end{abstract} 

The electronic properties of solids are closely related to their crystal structure which is strictly determined by the atomic nature of the individual elements. 
However, the imposition of a superstructure on a given lattice makes the controlled fabrication of structures with chosen properties possible. 
One approach discussed more recently is to place quantum dot (QD), also known as artificial atoms \cite{Ashoori,Tarucha}, on the points of a lattice to form an artificial crystal called a quantum-dot superlattice (QDSL).
Prerequisites for obtaining a band structure based on the periodic potential in a quantum-dot lattice are coherent coupling and good uniformity in the QD array. 
Recent experiments have reported the realization of such QDSLs. 
The observation of the quantum Hall effect in a lateral periodic potential based on a two-dimensional electron system \cite{Albrecht} is clear evidence of phase coherent transport across unit cells. 
Schedelbeck {\it et al.} \cite{Schedelbeck} used cleaved edge overgrowth to fabricate coupled QDs that formed in the intersection of three quantum wells. They observed peak-splitting in the photoluminescence spectrum demonstrating the coherent coupling of states localized in the QDs. 
The QDs formed because of the larger area at the intersections, which lowers the confining potential. In a similar way, QDSLs can be created using two-dimensional quantum wire networks which allow the formation of a large number of coupled dots with the high regularity of the underlying grid.
This type of wire network geometry has already been achieved with GaAs \cite{Kumakura} and Si \cite{Namatsu} wires.  

The use of QDSLs should make it possible to observe the electron correlation effects predicted in mathematical models of lattice systems, such as the Hubbard model. 
The Hubbard Hamiltonian is defined as $H = -\sum_{i,j,\sigma}{t_{ij}c_{i\sigma}^{+} c_{j\sigma}} + U \sum_{i} {n_{i\uparrow} n_{i\downarrow}}$, where $t_{ij}$ is the hopping integral between different sites $i$ and $j$, $U$ is the intra-site interaction between electrons with opposite spin, and $n_i$ ($c_i$) is the electron number (annihilation) operator with spin $\sigma$ on site $i$. 
Recently, we have used the spin dependent local density approximation (LDA) as a basis for proposing a feasible method 
for achieving a ferromagnetic ground state in a Kagome QDSL within a wide electron filling range \cite{Shiraishi}, 
which is consistent with a mathematical proof \cite{Mielke} 
of the ferromagnetism in the Hubbard model on the Kagome lattice. 

In this Letter, we propose a method for forming QDSLs in a quantum wire network of square and {\it plaquette} lattices. 
The plaquette lattice has a square plaquette in each unit cell with four lattice-points at the vertices, as shown in Fig. 1(b). 
It can be shown that both QDSLs are well represented by the Hubbard model.
An interesting difference between the two QDSLs is that the Fermi surface of the plaquette QDSL has disconnected pieces whereas the square QDSL is formed of one piece. 
To find a correlation effect that reflects both 
the Coulomb interaction and the structures of the Fermi surface, 
we studied the existence of superconductivity for both lattices within the framework of the Hubbard model. 
We found a superconducting ground state where the transition temperature $T_c$ of the plaquette lattice is more than double that of the square lattice and sufficiently high to allow superconductivity to be observed experimentally. 

We begin with the design of the square QD lattice. 
We consider the quantum wire network shown in Fig. 1(a). 
We assume InAs quantum wires buried in In$_{0.776}$Ga$_{0.224}$As 
barrier regions with a band offset of 0.17 eV \cite{Taguchi}. 
Each quantum wire is 50 nm wide and 50 nm high, 
and the lateral distance between adjacent wires is 61.1 nm. 
These parameters have been carefully determined. 
If the lateral distance is too small, the wave function of an electron 
is extended and QDs are not appropriately formed. 
By contrast, if the distance is too large, the wave function is 
too localized at the intersection and the hopping parameter between adjacent QDs becomes very small. 
The electronic band structures are obtained 
using first-principles calculations based on the LDA \cite{Shiraishi} 
with the Perdew-Zunger exchange-correlation potential \cite{Perdew}.
We assume the effective mass of electrons to be $m^*=0.02m_0$ 
and the dielectric constant
$\epsilon=12.4\epsilon_0$ for the InAs wires. We employ conventional 
plane wave expansion for the LDA calculation, and include a sufficiently large number of plane waves to ensure numerical convergence. 

The LDA band diagram for this wire network is shown in Fig. 2. It can be perfectly fitted to the tight-binding calculation for the square lattice parametrized in Fig. 1(b): the fitting parameters are hopping integrals between adjacent sites ($t_a=t_b=0.179$ meV) 
and much smaller ones between non-adjacent sites 
($t_c=t_d=t_f=-0.017$ meV and $t_e=-0.021$ meV). 
This fitting result indicates that the tight-binding approximation 
is very accurate and the wave function is appropriately localized on a QD \cite{note1}.
The localization originates from the coherence between 
the crossed plane waves at the intersection, resulting in a  
localized state with reduced energy. 
From another point of view, because of the uncertainty condition 
between position and momentum in quantum mechanics, 
the momentum of electrons, or the kinetic energy, 
becomes lower at a wider area in the intersection. 
Note that the results for fitted hopping parameters 
between the nearest neighbor QDs are sufficiently large and the wave function is not too localized on a QD. 
We should also note 
that the fitted hopping parameters between the second nearest 
dots ($t_c$, $t_d$, and $t_f$) and those between the third nearest dots 
($t_e$) are small. 
This is due to the large band-offset of In$_{0.776}$Ga$_{0.224}$As which suppresses tunneling through this region. 
Therefore, the hopping parameters are determined by 
the distance not in a straight line but along the wires. 
The localization of the wave function also suggests 
that the intra-dot Coulomb interaction 
between electrons is large. 
The intra-dot interaction, simply estimated by subtracting 
the Ewald sum for inter-dot interactions from the total Hartree energy in the LDA, 
is $U=1.8$ meV, which is similar to that estimated in a QD formed 
in a spherically symmetric harmonic potential 
with a 50 nm oscillator length. 
The value of $U/t_a\sim 10$ shows 
that the electrons are strongly correlated 
and the many-body effects of the electrons should be pronounced. 

Next, we consider the plaquette QDSL [Fig. 1(a)], which we obtained by alternating the distance between the wires in the wire network for the square QDSL. 
In our design, the smaller (larger) distance 
between adjacent wires is 38.8 (83.4) nm. 
In the tight-binding model parametrized in Fig. 1(b), the hopping parameter between adjacent sites within a plaquette and the one across two plaquettes are different ($t_a\ne t_b$) unlike in the square lattice ($t_a=t_b$). 
By performing the same analysis undertaken for the square lattice, 
we confirmed that the tight-binding model 
provides a good fit with the LDA band diagram 
with the fitting parameters $t_a=0.242$, $t_b=0.151$, $t_c=0.003$, 
$t_d=-0.025$, $t_e=-0.024$, and $t_f=-0.014$ in milli-electron volt units. The estimated intra-dot interaction is 1.7 meV. 

A characteristic difference between the square QDSL 
and the plaquette QDSL lies in the structure of the Fermi surface shown 
in Fig. 3, which is calculated in the LDA with the electron filling 
$n=0.8$ (=number of electrons/number of QDs). 
The Fermi surface of the square QDSL is connected, 
while that of the plaquette QDSL has disconnected pieces 
originating from the band folding. 
To demonstrate that the difference in the structure of the Fermi surface significantly influences the characteristics of the correlated electrons in QDSLs, we studied superconducting correlations in both QDSLs
by assuming the Hubbard model with the estimated hopping 
and intra-site interaction parameters described above. 
It has been shown that the repulsive Hubbard model exhibits a superconducting transition.
In addition to early calculations \cite{Scalapino}, 
a recent quantum Monte Carlo calculation \cite{Kuroki&Aoki} 
indeed shows the enhancement of the pairing 
correlation with $d$-wave symmetry in the two-dimensional repulsive Hubbard model. 
Here, we employ the fluctuation-exchange (FLEX) approximation \cite{Bickers,Bickers&White,Pao,Monthoux,Grabowski,Dahm}
with the Eliashberg equation \cite{Eliashberg} 
in order to evaluate the superconducting transition temperature $T_c$ in the Hubbard model 
for the plaquette and square lattices. 
The FLEX approximation is a kind of self-consistent random phase approximation 
and is known to be appropriate for treating the strong antiferromagnetic 
spin fluctuation that originates from the Coulomb interaction  
between electrons with opposite spin.
We note that the FLEX approximation gives a good estimate of $T_c$ 
for cuprates \cite{Bickers,Bickers&White,Pao,Monthoux,Grabowski}.
For both lattices, we assume the filling of electrons $n=0.8$, where the system is far from the antiferromagnetic instability region near half-filling ($n=1$) but the spin fluctuation is still large enough to obtain a high $T_c$. 
Our analysis shows that $T_c$ for the plaquette QDSL is 90 mK, which is more than double that of $T_c$=40 mK for the square QDSL. 
Both $T_c$ values are high enough for us to observe superconductivity experimentally. 
However, low-temperature measurements are usually limited to 
a few tens of milli-Kelvins, and a higher $T_c$ would be 
preferable as it would allow us to observe the superconducting transition more clearly. 
Figure 4 shows the improvement in $T_c$ when the hopping parameters $t_a$ and $t_b$ 
are changed for $n=0.85$ and $U=7$. Other small hopping parameters are disregarded 
($t_c=t_d=t_e=t_f=0$) because they have little influence on the result. 
Energy is scaled in units of average hopping energy
$t=(t_a+t_b)/2=1$. In these units, the single-particle bandwidth 
is simply given by $4(t_a+t_b)=8$. 
We find that $T_c$ is enhanced as the difference 
between $t_a$ and $t_b$ is increased, 
and $T_c$ for $t_a=1.5$ and $t_b=0.5$ is more than double that for the square lattice ($t_a=t_b=1$). 
 
The mechanism of this large $T_c$ enhancement is similar to 
that recently proposed by Kuroki and Arita \cite{Kuroki&Arita}. 
The strength of the superconducting pairing interaction is characterized by the coupling constant in the form
\begin{eqnarray}
V_{\rm eff}=-\frac{\sum_{\vec{k},\vec{k}^\prime\in {\rm Fermi\ surfaces}}
V_{\rm pair}(\vec{k}-\vec{k}^\prime)\phi(\vec{k})\phi(\vec{k}^\prime)}
{\sum_{\vec{k}\in {\rm Fermi\ surfaces}}\phi(\vec{k})^2}
\end{eqnarray}
where $V_{\rm pair}$ is the effective electron-electron interaction induced 
by the pair-scattering process from $[\vec{k},-\vec{k}]$ to $[\vec{k}^\prime,-\vec{k}^\prime]$ on the Fermi surface 
and $\phi(\vec{k})$ is the superconducting gap function.
Positive values of $V_{\rm eff}$ 
 correspond to an attractive interaction and, 
roughly speaking, a larger $V_{\rm eff}$  gives a higher $T_c$. 
Reflecting the Fermi surface nesting in the Hubbard model, antiferromagnetic spin fluctuations near half filling 
enhance the spin susceptibility  $\chi(\vec{q})$ 
[roughly proportional to $V_{\rm pair}(\vec{q})$] 
around a certain wave vector  $\vec{q}=\vec{Q}$.
When pair scattering on the Fermi surface is accompanied by a momentum transfer  
$\vec{k}-\vec{k}^\prime$ around the peak $\vec{Q}$ 
and a sign change of the gap function $\phi(\vec{k})\phi(\vec{k}^\prime)<0$, 
this scattering process gives significant positive contributions to $V_{\rm eff}$. 
In the square lattice, the Fermi surface is perfectly nested 
at half-filling and  $\chi(\vec{q})$ 
exhibits the antiferromagnetic peak at $\vec{Q}=(\pi,\pi)$. 
Since the 
gap function $\phi(\vec{k})=\cos k_x-\cos k_y$ with $d_{x^2-y^2}$ symmetry
 has nodes on the Fermi surface, some pair scatterings transferring 
$\vec{k}-\vec{k}^\prime=\vec{Q}\approx (\pi,\pi)$ 
between two points [indicated by solid arrows in Fig. 3(a)], 
where the gap function has the opposite sign, 
make significant positive contributions to $V_{\rm eff}$. 
However, some other scatterings 
$\vec{k}-\vec{k}^\prime=\vec{Q}+\Delta\vec{Q}$ with the same sign of $\phi(\vec{k})$
(indicated by dotted arrows) reduce $V_{\rm eff}$. In the plaquette lattice, 
the Fermi surface is folded in half and an antiferromagnetic peak emerges 
at $\vec{Q}\approx0$ 
with a finite spread $\Delta\vec{Q}$. 
The Fermi surface is separated by lines $|k_x|=|k_y|$, 
and is separated into pieces, each of which has a gap function 
with the same sign, as shown in Fig. 3(b). 
Dominant scattering processes around $\vec{Q}\approx0$ 
always make positive contributions 
to $V_{\rm eff}$ and significantly enhance $T_c$. 
For a dimerized lattice, 
which is somewhat similar to a ladder system \cite{Dagotto}, 
Kuroki and Arita \cite{Kuroki&Arita} have shown that a very high $T_c$ can be reached based on this mechanism. 
However, to raise $T_c$ in a dimerized lattice, 
it is necessary to tune three different hopping-parameters, whereas, in plaquette lattice-patterns, only two hopping parameters determine $T_c$, and these can be easily modulated by selecting the wire geometry. 

In summary, we proposed a QDSL design based on square and plaquette type structures in quantum wire networks. 
Both QDSLs are well represented by the Hubbard model. 
We studied superconducting transitions for both QDSLs using the Hubbard model and found the $T_c$ for the plaquette QDSL 
to be more than double that for the square QDSL. 
The estimated $T_c$ value in the plaquette QDSL is sufficiently high to allow us to observe superconductivity experimentally. 
Finally, we note that the transition temperature of cuprates might be increased close to room temperature if the material is patterned in a plaquette type lattice.

The authors are grateful to K. Sakairi for useful discussions,  
and S. Kurihara and H. Aoki for stimulating discussions. 
They also thank A. Richter for his critical reading of the manuscript and S. Ishihara for his continuous encouragement 
and helpful advice. This work was partly supported by the NEDO 
International Joint Research Grant, 
the NEDO under the Nanotechnology Materials Program, and ACT-JST.


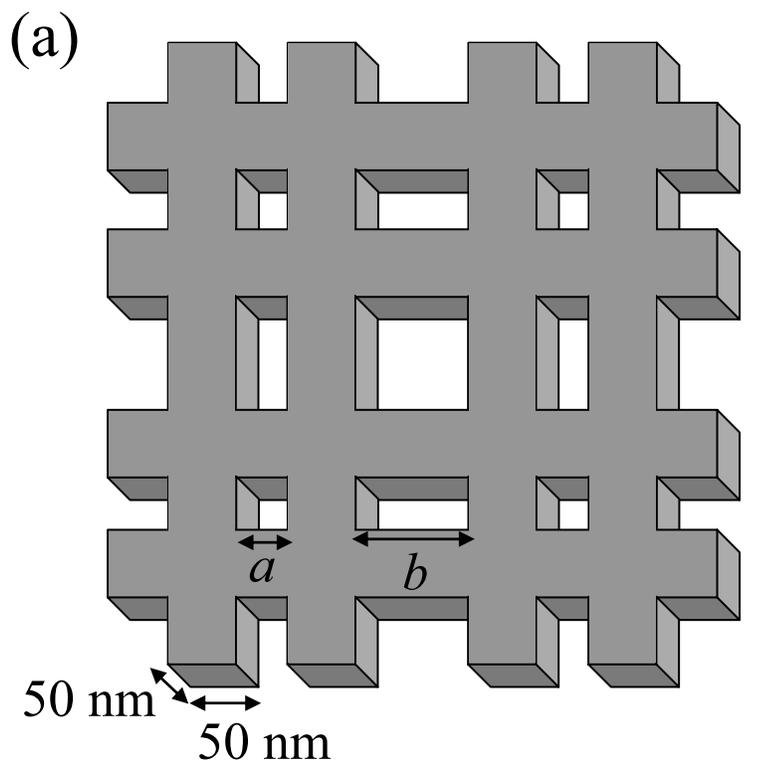
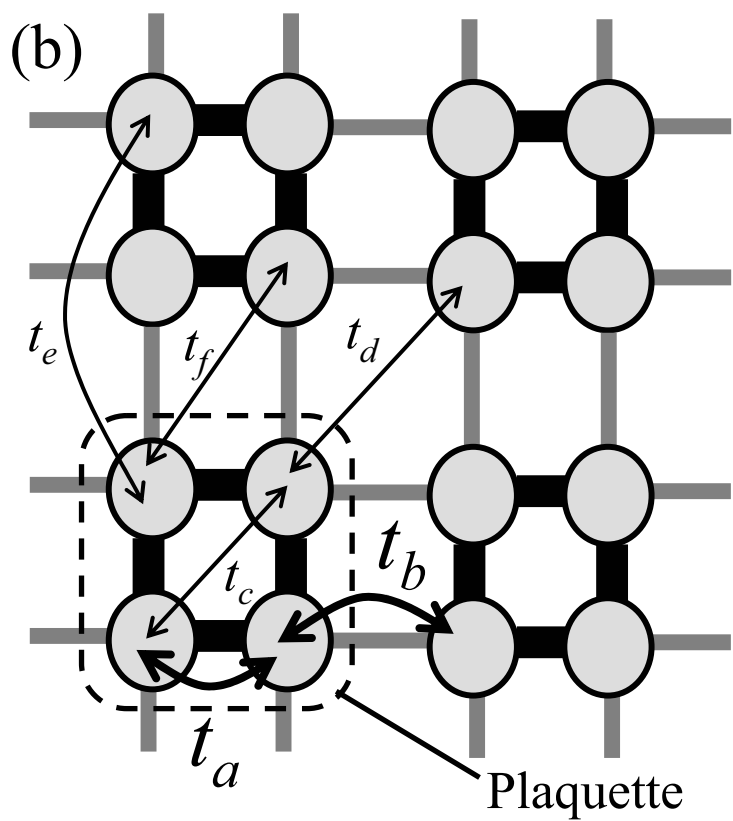
\begin{figure}
\caption{
(a) Diagram of a nonuniform quantum wire network with distances $a$ and $b$ between 
adjacent wires. 
(b) Corresponding tight-binding model with hopping parameters $t$. 
For a square QDSL, $t_a=t_b$ and $t_c=t_d=t_f$. 
}
\label{fig1}
\end{figure}

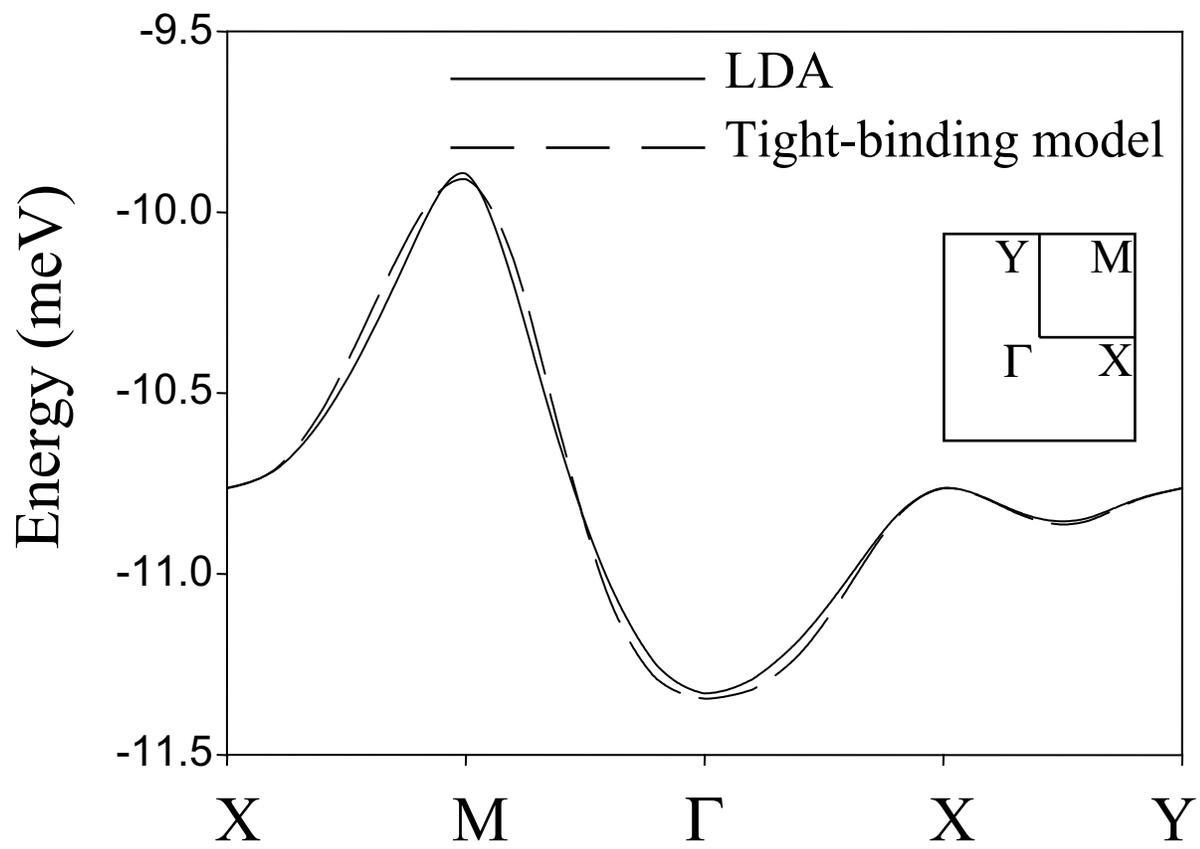
\begin{figure}
\caption{
Result of fitting the band structures of the square QDSL. 
The LDA band diagram of a quantum wire network (solid curve) 
and the best fit in the tight-binding calculation 
(dashed line) with parameters $t_a=t_b=0.179$ meV, 
$t_c=t_d=t_f=-0.017$ meV, and $t_e=-0.021$ meV are shown. 
}
\label{fig2}
\end{figure}

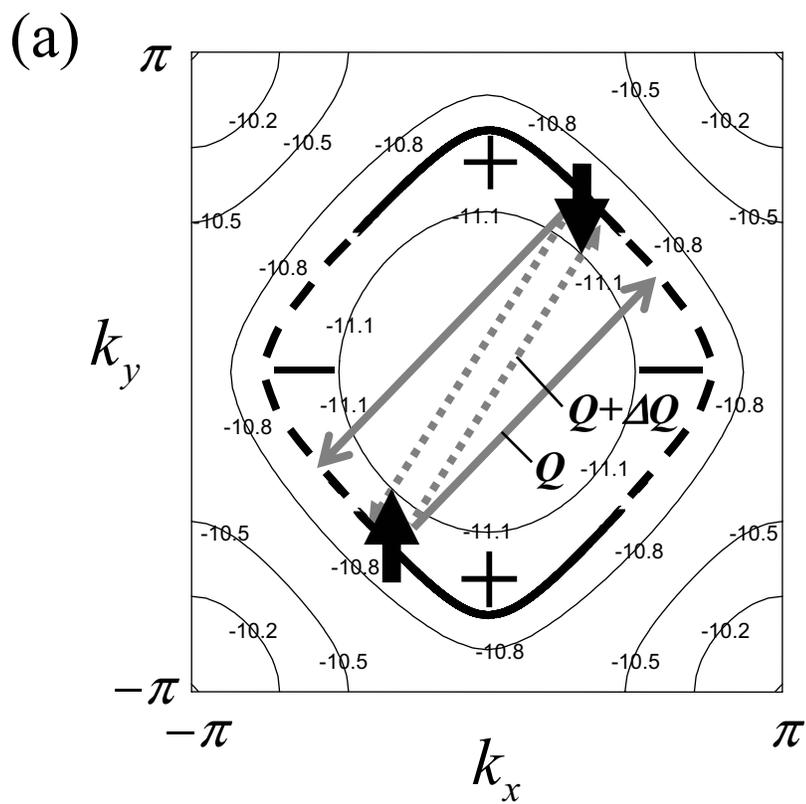
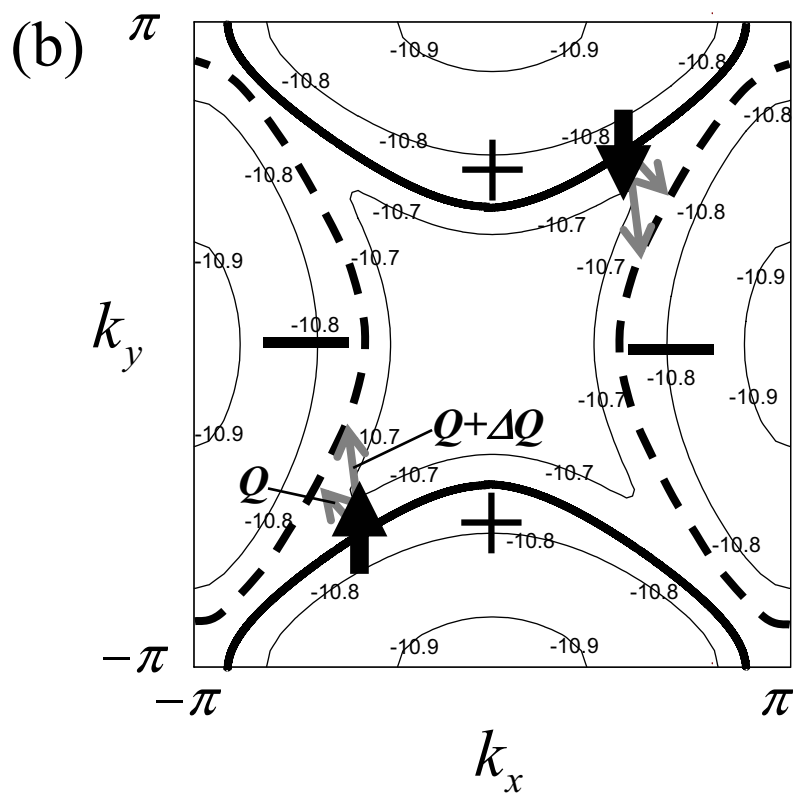
\begin{figure}
\caption{
Energy contours for (a) the square QDSL and (b) the plaquette 
QDSL calculated in LDA. The energies are plotted in milli-electron volts. 
The thick solid (dashed) curves represent the Fermi surfaces 
where the superconducting gap function is positive (negative). 
The upward (downward) pointing arrow indicates 
an electron with up (down) spin. 
The solid (dashed) arrow schematically represents 
dominant pair-scattering processes that give a positive 
(negative) contribution to the pairing interaction by transferring a momentum 
at which the spin susceptibility has a peak with a finite spread. 
}
\label{fig3}
\end{figure}

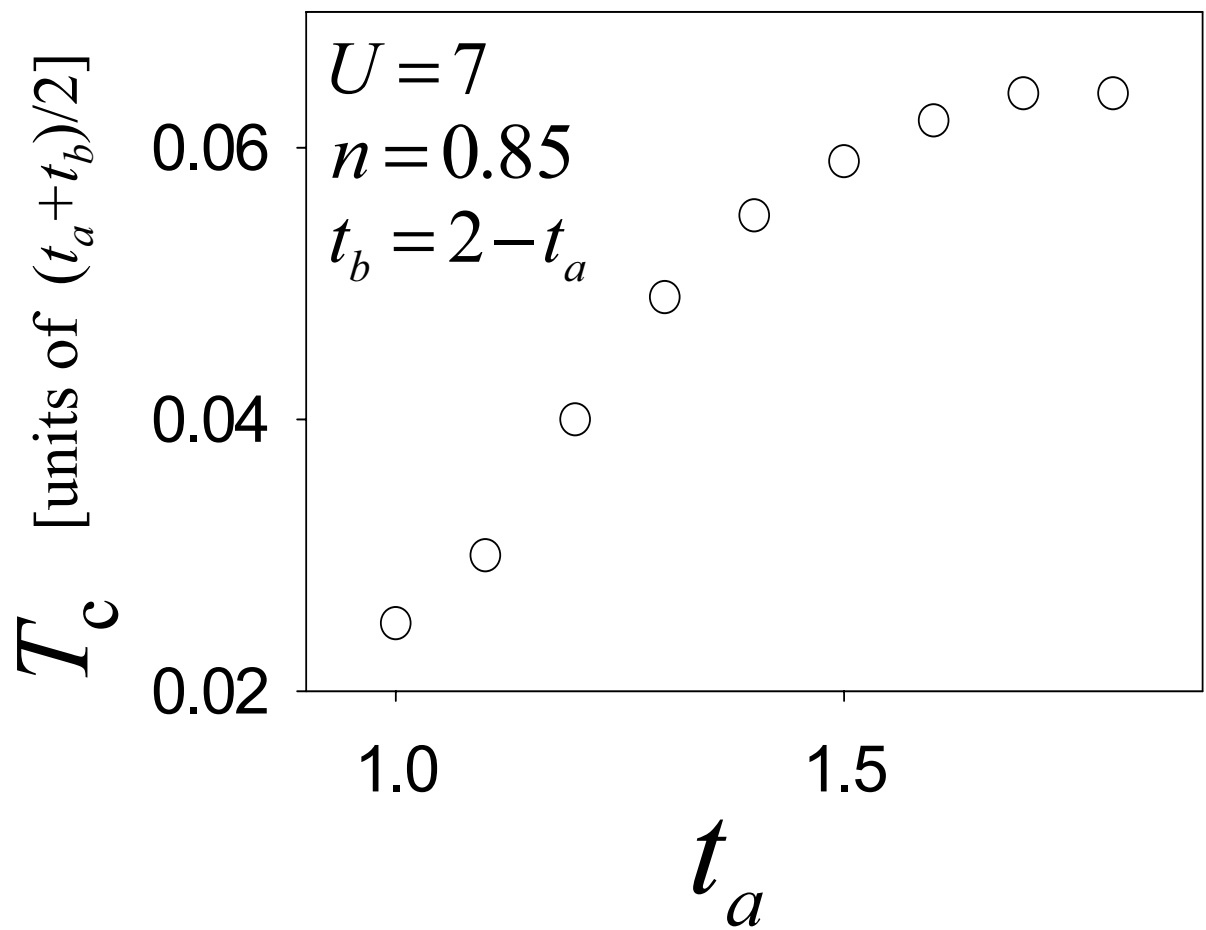
\begin{figure}
\caption{
$T_c$ plotted as a function of $t_a$ in units of the 
average hopping parameter $t=(t_a+t_b)/2=1$.  
The single-particle bandwidth is given by $4(t_a+t_b)=8$. 
Parameters $t_a=t_b=1$ (for the square lattice) gives $T_c=0.025$. 
}
\label{fig4}
\end{figure}


\begin{references}
\bibitem{note1} We have confirmed that the wave function is indeed localized in the intersection of two crossed quantum-wires by solving the Schr{\"o}dinger equation numerically.
\end{references}
\end{document}